Title: **A hadron model with breaking of spatial homogeneity of vacuum.**
Authors: **E.N. Myasnikov, E.V. Mastropas (Southern Federal University)**

A possible breaking of spatial homogeneity of vacuum due to the interaction between quark and Bose-field is analyzed. It is shown that in this case quark can be in a localized state (like wave packet). Energetic conditions for such a spontaneous symmetry breaking are found in suggested model. Possible consequences of such symmetry breaking, in particular, the origin of deep inelastic processes and quark confinement phenomenon are discussed.

The importance of spontaneous symmetry breaking phenomenon in modern elementary particle physics and quantum field theory can hardly be overestimated [1, 2]. And we find it necessary to consider all known symmetries from this point of view. We consider it promising to analyze possible consequences of rejecting a concept of spatially homogeneous vacuum in the neighbourhood of a hadron. Indeed, if each quark within a hadron deforms vacuum, it would lead to construction of a classical field, which would accompany quark in its motion and be centered within a wave packet. And hadron obviously acquires a number of observable features whose origins are explained now by using of other models.

In particular, if the area of quark localization is rather small in such a wave packet, then none of hadrons can have a symmetry center. A hadron state will not correspond to any certain value of spatial parity with all obvious consequences for reactions and for decays. Similarly, its states will not also be CP-invariant [3]. If all the quarks in hadron deform vacuum then away from hadron vacuum will be different from that inside of a hadron. Who knows, may be solution to the quark confinement problem [4] can be found this way? A baryon in a ground state can consist of three equal quarks if each quark inside a hadron is in the state of rather narrow wave packet. In this case quark's states can differ by the localization area inside the baryon instead of common practice to distinguish quarks based on the color charges.

We would like to point out that the "spontaneous symmetry breaking" term is not really accurate. In fact, spontaneous symmetry breaking is caused by interaction between the two of quantized fields. The occurrence of states with vacuum deformation is energetically beneficial even if operator of interaction $H_i$ is symmetric. Assume $H_i$ describes, for example, processes with radiation and absorption of some Bose-field quantums by quark (with preservation of momentum in such processes). If the interaction is strong enough and if each boson has the rest mass $m$, a field operator of these bosons will get nonzero average values in the area of quark wave packet localization. In other words, we have a Bose-condensate arising in Bose-field as an element of vacuum. Below we will show that in this case existence of this condensate is energetically favorable as it shields a charge of quark, which Bose-field interacts with. Below we will show that in the case of strong enough interaction between the quark and a massive vector Bose-field quark can be in a localized state and Bose-condensate of this field will exist in the area of quark localization. A quark mass defect in this state appears to be about mass of the non-localized quark.

Clearly, if quark interacts with several Bose-fields, then in the area of its localization Bose-condensates of these fields would appear as a result of quark's transition into the state similar to wave packet. Screening of all the quark charges which occurs during such a process will result in decrease of its interaction with other quarks. This observed effect has received the name of "asymptotic freedom". Investigation of a possibility of spatially inhomogeneous Bose-field's condensate construction seems to be of obvious importance.

For this purpose let us consider a simplest model, which consists of the quark and Bose-field interacting with quark. Assume this interaction keeps a quark flavor similar to interaction between the quark and intermediate Z-bosons [1, 2]. Therefore bosons under consideration we further will call "Z-bosons", although boson mass $m$, a mass $M$ of a "bare" quark and interaction constant $a$ will be varied in system of units ($\hbar = c = 1$). We'll be looking for a binding energy $M - M_0$ (where $M_0$ is energy of quark in a localized state) and localization parameter $r_0$, which will be introduced below.

In order to simplify the method for solving a problem let us consider a quark state as a rest wave packet. It is rather easy to pass to the solution for moving wave packet using Lorentz transformation. The method of calculation used below can be applied if a root-mean-square of a quark momentum $p$ in the state considered will be less than its "bare" mass $M$ and if the boson mass will be less than quark's mass defect $M - M_0$.

In such a simplest model only a time component of quark current included into the interaction operator will have a nonzero value. Square module of the only nonzero component $\psi(\vec{r})$ of Dirac wave function [5] of quark will be proportional to this component. As a vector of a state of our system we use

$$\psi |0\rangle, \qquad (1)$$

where $|0\rangle$ is a vector of ground state of Z-bosons field.

To describe Z-bosons condensate it is convenient to separate out the classical variables. Using a unitary operator [6]

$$U = \prod_{\vec{k}} \exp\{d_{\vec{k}} b_{\vec{k}}^+ - d_{\vec{k}}^* b_{\vec{k}}\} \qquad (2)$$

(where $b_{\vec{k}}^+$ and $b_{\vec{k}}$ are operators of creation and annihilation of Z-bosons with momentum $\vec{k}$), we will transform Hamiltonian of a system and a vector of system state (1).

In the state $U|0\rangle$ an average value of $b_{\vec{k}}$ operator will be nonzero and will be equal to $d_{\vec{k}}$, i.e. in this state an average value of Z-bosons field operator will be nonzero. Hence a Z-boson condensate will exist, which form will be defined by $d_{\vec{k}}$ amplitudes. Transformation (2) in Hamiltonian replaces birth and annihilation operators by rules $b_{\vec{k}} \to b'_{\vec{k}} + d_{\vec{k}}$, $b_{\vec{k}}^+ \to b'^{+}_{\vec{k}} + d_{\vec{k}}^*$, and new operators $b'_{\vec{k}}$ and $b'^{+}_{\vec{k}}$ have zero average values in $U|0\rangle$ state.

Naturally, in $U|0\rangle$ state the average values of $n_{\vec{k}} = b'^{+}_{\vec{k}} b'_{\vec{k}}$ operators are equal to zero as well. Therefore while calculating the average value of transformed Hamiltonian $H' = UHU^{-1}$ in $\psi U|0\rangle$ state it is possible to neglect all the terms that contain $b'_{\vec{k}}$ and $b'^{+}_{\vec{k}}$. $H'$ will be transformed after this operation into a function of $d_{\vec{k}}$ and $d_{\vec{k}}^*$ parameters, as well as field operator $Z$ will be transformed into a function of spatial coordinates. For example, in order to calculate the average value it is possible to represent the $H_i$ operator as a functional

$$H_i = a \int d\vec{r}\, Z_0(\vec{r}) |\psi(\vec{r})|^2 \qquad (3)$$

containing only the time component $Z_0(\vec{r})$ of bosons field function.

In a considered case of motionless localized quark a Lagrange equation will look like

$$(\Delta - m^2) Z_0 = a|\psi|^2. \qquad (4)$$

At a $\delta$-function right part the solution of equation (4) is Yukawa potential with a Fourier-image

$$Z'_0(\vec{k}) = a(m^2 + k^2)^{-1}. \qquad (5)$$

The solution of equation (4) can be presented as

$$Z_0(\vec{r}) = a \int d\vec{k}\, (2\pi)^{-3} (m^2 + k^2)^{-1} |\psi|^2_{\vec{k}} \exp(i\vec{k}\vec{r}), \qquad (6)$$

where $|\psi|^2_{\vec{k}}$ is a Fourier-image of $|\psi(\vec{r})|^2$ function.

Let us introduce parameter $r_0$ of localization by defining a factor in expansion of the wave function $\psi(\vec{r})$ in terms of wave functions of states with certain momentum as

$$\psi_{\vec{k}} = \frac{8}{\pi} \sqrt{\frac{2r_0^2}{7}} \left(1 + k^2 r_0^2\right)^{-3}. \tag{7}$$

According to (7), the Fourier-image of $|\psi|_{\vec{k}}^2$ depends on $r_0$ as follows:

$$|\psi|_{\vec{k}}^2 = \frac{64}{7} \frac{28 + r_0^2 k^2}{\left(4 + r_0^2 k^2\right)^4}. \tag{8}$$

It is not difficult to prove that for the function $Z_0$ satisfying equation (4), Hamilton function $H_Z$ of a free Z-bosons field will be connected to $H_i$ by the following relation:

$$H_Z = -\frac{1}{2} H_i. \tag{9}$$

Hence quark energy in localized state goes down due to interaction with Z-field condensate at any quark localization radius $r_0$. An average value of a system Hamiltonian $H$ in state (1) can be represented as follows:

$$H = \int d\vec{k} \sqrt{M^2 + k^2} \, \psi_{\vec{k}}^2 + \frac{1}{2} H_i, \tag{10}$$

$$H_i = -\frac{a^2}{2} \int d\vec{k} \left(m^2 + k^2\right)^{-1} \left(|\psi|_{\vec{k}}^2\right)^2. \tag{11}$$

These integrals depend on parameter of localization $r_0$. We can try to select wave function $\psi$ in such a manner that this average value will be close to the ground state energy of the system. We have done this selection by varying parameter $r_0$. Setting a derivative of $H$ by $r_0$ to zero, let's find a minimum, which would depend on $M$, $m$ and $a$. It appears that at $a > a_c$ ($a_c = 4.135$) there's no minimum, and the system tends to collapse with $r_0 \to 0$. At $a \leq a_c$ there is one minimum, which position and depth depend on $M/m$. At $a = a_c$ this minimum corresponds to $r_0 = 0$, and mass defect practically coincides with $M$. But this case falls outside the limits applicability of our theory. As $a$ decreases, minimum displaces to the area of bigger $r_0$, and does it the slower the higher $M/m$ is. The depth of this minimum $M - M_0$ (i.e. the mass defect) thus decreases rapidly.

For example, at $a = 4$ and $\frac{M}{m} = 5$ relative defect of quark mass $\frac{M - M_0}{M}$ due to connection with bosons condensates up to 40% while the localization radius $r_{0\min}$ is close to $(10m)^{-1}$ (i.e. 10 times less than Compton wave length of a boson). A root-mean-square momentum of quark in this state ranges up to 60% of $M$. In this case, all applicability conditions of approximations used by us are hold true. If we set $m = 10\,MeV$, then radius of quark localization appears to be about baryon radius and hence baryon state will not be P- and CP-invariant. A mass of localized quark appears to be equal to 30 $MeV$.

We emphasize that this model can be improved. Undoubtedly, the increase in mass defect and decrease in localization radius are possible due to optimization of a wave function $\psi$ using several variational parameters. It is possible to cancel the restriction of $p < M$, and take into account contribution to the mass defect of interaction between quark and other massive Bose fields, which would also screen quark charge and will result in increase of effective coupling constant. Apparently, their mutual influence can increase essentially mass defect of three quarks in baryon.

In the further perspective we can consider some optimistical case of a baryon structure. Let us assume that quark localization occurs under the following conditions: $a \approx 1$, $\frac{M}{m} \approx 10^4$ and $m \approx 1 MэB$. It can be that quark mass defect is $M - M_0 \approx 9990 \, MeV$ and mass of localized quark is 10 $MeV$ with radius of localization equal to $r_0 = (1000m)^{-1}$. In this case which is beyond the scope of the theory described above, let's consider collision of a fast lepton with quark from inside of a hadron. Let us introduce $A$ as minimal energy loss necessary for break off connections between quark and vacuum deformation. Then under the law of preservation we'll receive (in the case $M_0 << M$): $A = E_{def} + M$, where $E_{def}$ is energy of deformed vacuum. As $M_0 = E_{def} + M + T + E_{int}$, where $E_{int} = -2E_{def}$ according to shown above, and as average kinetic energy $T$ in considered case $M_0 << M$ will be more than $M$ ($T \to \infty$ with $r_0 \to 0$), then $E_{def} > 2M$ and $A > 3M$. The minimal value of $A$ should be equal to $40 \, GeV$ for the considered case with $M = 10 \, GeV$.

And if Bose-condensates of all three quarks in baryon are interconnected, the energy necessary for quark to break off connections with vacuum deformation will approximately increase by 80 GeV. Most likely, in such complex system as baryon, the result of a collision will be intensive radiation of the mesons arising in attempt of quarks to skip through such a high energetic barrier. So in this case deep inelastic process and quark confinement will be observed. We believe that our explanation of the confinement phenomenon offer obvious advantages over the presently accepted model according to which the energy of interaction between quarks grows with a growth of distance between them. It can be that further investigation of our theory will allow us to find out some possible quasistationary values of $M_0$ at big $M$. And it will allow us to understand observable strong dependence of $M_0$ on a quark flavor.